\documentclass[twocolumn,aps,prb,showpacs,preprintnumbers,amsmath,amssymb,superscriptaddress]{revtex4}

\usepackage{graphicx}
\usepackage{dcolumn}
\usepackage{bm}
\usepackage{color}

\begin{document}


\title{Phonon Density of States and Thermodynamic Behavior in Highly Amorphous Media}


\author{D. J. Priour, Jr}
\affiliation{Department of Physics, University of Missouri, Kansas City, Missouri 64110, USA}


\date{\today}

\begin{abstract}
We calculate the phonon density of states (DOS) for strongly amorphous 
materials with a short-ranged interatomic potential.  Exponentially
decaying and abruptly truncated interatomic potentials are examined.
Thermally excited mean square deviations from equilibrium are calculated
with rapid increases noted as the average number of neighbors 
is reduced.
The Inverse Participation Ratio (IPR) is used to characterize the 
phonon states and identify localized phonon modes as the bonding range
(and hence the average number of neighbors per atom) is diminished.
For the truncated potential, the characteristics of the IPR histogram
change qualitatively below $n_{\mathrm{neigh}}$ with the appearance of  
localized phonon modes below $n_{\mathrm{neigh}} = 6.0$.
\end{abstract}

\maketitle


\section{Introduction and methods of calculation}
The range of materials used in current engineering applications is diverse,
and many structural media have an amorphous character.  
Acoustic and thermodynamic properties relevant to the performance of a class of 
materials are technologically significant properties. 
We examine highly amorphous systems and calculate the phonon frequency density  
of states (DOS) which may be used to determine the thermally 
excited deviations from equilibrium.
In addition, the Inverse Participation Ratio (IPR),
$\kappa_{\mathrm{IPR}} = 
\langle \lvert \psi \rvert ^{4} \rangle/\langle \lvert \psi \rvert ^{2} \rangle^{2}$, 
provides a way to characterize the phonon modes and determine the degree to which they are 
localized~\cite{uno,dos}.  In the moment ratio $\kappa_{\mathrm{IPR}}$, $\psi$ is the positional amplitude of a phonon state, 
and vibrational modes with localized character are associated with a finite value of 
$\kappa_{\mathrm{IPR}}$ whereas for extended states $\kappa_{\mathrm{IPR}}$ tends to 
zero in the bulk limit.

An important question is the effect of changing the range of the coupling between 
atomic species.  A short-ranged coupling corresponding to a relatively loosely 
packed system will lead to reduced rigidity, while a longer ranged coupling 
increases the average number of nearest neighbors, and thereby imparts rigidity to the 
amorphous lattice.

We consider a coupling scheme with an abrupt cutoff $r_{c}$, and we also examine an exponentially 
decaying interaction scaling as $e^{-\lambda r}$ where the rapidity of the decay may be tuned with a finite 
(though rapidly diminishing) coupling as a function of distance by changing 
the damping constant $\lambda$.  Hence, with genuine
discontinuities absent from the interaction function, qualitatively similar (albeit less
abrupt) changes will occur as the parameter controlling the rapidity of decay is 
modified.
The flat potential terminating beyond a threshold radius $r_{c}$ is examined in Section II, 
and in Section III results are discussed for the exponential coupling scheme.

In analyzing lattice dynamics, we use the classical Lagrangian formalism 
\begin{align}
\mathcal{L} = T - V = \frac{1}{2} \sum_{i=1}^{N} m_{i} \dot{x}_{i}^{2} + \frac{1}{2}     
\sum_{i=1}^{N} \sum_{j=1}^{n_{i}} V_{ij} (r_{ij}),
\end{align}
where the factor of $1/2$ in the potential is included to compensate for 
redundant counting of bond energies, $n_{i}$ specifies the number of neighbors 
for the atom given the label $i$, and $V_{ij}(r)$ is the specific interaction 
between atoms $i$ and $j$.
In the context of our calculation, the system is contained in a supercell with cubic geometry and length $L$, 
containing $N$ atoms of equal mass $m$ with periodic boundary conditions 
assumed.    

In the temperature regimes of interest, we assume the deviations in atomic positions to 
be small in relation to the equilibrium bond lengths $l_{ij}^{0}$, and 
we employ the harmonic approximation (i.e. essentially a Taylor expansion of a 
well behaved potential) with 
\begin{align}
V = \frac{1}{2} \sum_{i=1}^{N} \sum_{j=1}^{n_{i}} \frac{K_{ij}}{2} ( l_{ij} - l_{ij}^{0})^{2},
\end{align}
with $K_{ij}$ being the second derivative of $V(r_{ij})$ at $l_{ij}^{0}$.  
In spite of the implementation of the harmonic approximation in the inter-atomic bonding 
potential, the non-collinearity of bonds will still give rise to anharmonic terms in the potential.

Nevertheless, a further harmonic approximation may be appropriate if the atomic coordinates are 
written in terms of the equilibrium coordinate with a shift term, such as $x_{i} = x_{i}^{0} + 
\delta_{i}^{x}$ with similar expressions for the $y$ and $z$ coordinates; the bond length $l_{ij}$ 
has the form 
\begin{align} 
l_{ij} = \sqrt{\begin{array}{l} (\Delta_{ij}^{0x} + \delta_{i}^{x} - \delta_{j}^{x})^{2} + 
(\Delta_{ij}^{0y} + \delta_{i}^{y} - \delta_{j}^{y})^{2} \\ + 
( \Delta_{ij}^{0z} + \delta_{i}^{z} - \delta_{j}^{z})^{2} \end{array}},
\end{align}
where $\Delta_{ij}^{0x} \equiv (x_{i}^{0} - x_{j}^{0} )$, 
$\Delta_{ij}^{0y} \equiv (y_{i}^{0} - y_{j}^{0} )$, and 
$\Delta_{ij}^{0z} \equiv (z_{i}^{0} - z_{j}^{0} )$.  Hence, the 
potential energy stored in the strongly disordered lattice depends 
only on the difference of coordinates such as, e.g., $\Delta_{ij}^{0x}$ for 
the equilibrium $x$ coordinates and $( \delta_{i}^{x} - \delta_{j}^{x} )$ for the corresponding 
shifts from equilibrium.  If the latter are sufficiently small relative to the former,
it is appropriate to expand about $\Delta_{ij}^{0x}$, $\Delta_{ij}^{0y}$, and $\Delta_{ij}^{0z}$, yielding
$(l_{ij} - l_{ij}^{0})^{2} \approx \left[ \hat{\Delta}_{ij} \cdot ( \vec{\delta}_{i} - \vec{\delta}_{j} ) \right ]^{2}$ where
$\hat{\Delta}_{ij}$ is a unit vector formed from the difference $\vec{\Delta}_{ij}^{0} = \mathbf{x}_{i}^{0} - 
\mathbf{x}_{j}^{0}$.  The interatomic potential, to quadratic order in components of $\vec{\delta}_{i}$ and 
$\vec{\delta}_{j}$ is 
\begin{align}
V_{\mathrm{Har}} = \frac{1}{2} \sum_{i=1}^{N} \sum_{j=1}^{n_{i}} \frac{K_{ij}}{2} \left[ 
\hat{\Delta}_{ij} \cdot (\vec{\delta}_{i} - \vec{\delta}_{j} ) \right ]^{2} 
\end{align}
The standard Lagrangian formalism then yields the equations of motion, given by, e.g. 
$\tfrac{\partial}{\partial t} \tfrac{\partial \mathcal{L}}{\partial \dot{\delta}_{i}^{x}} 
+ \tfrac{\partial \mathcal{L}}{\partial \delta_{i}^{x}}  = 0$.
Written in terms of the vector displacement $\vec{\delta}_{i}$, one has
(we assume the atoms to have identical mass $m$)
\begin{align}
\frac{d^{2}}{dt^{2}} \vec{\delta}_{i} = -\sum_{j=1}^{n_{i}} \frac{K_{ij}}{m} \left [ \hat{\Delta}_{ij} \cdot 
(\vec{\delta}_{i} - \vec{\delta}_{j}) \right]\hat{\Delta}_{ij}
\end{align} 
Using the ansatz $\vec{\delta_{i}} = \vec{\delta_{i}}^{\mathrm{A}} e^{i \omega t}$ 
removes the explicit time dependence and reduces the solution of the 
equations of motion to an eigenvalue problem, which may be written as
\begin{align}
\omega^{2} \! \left[ \! \! \begin{array}{l} \delta_{i}^{x} \\ \delta_{i}^{y} \\ \delta_{i}^{z} \end{array} \! \! \right] \! = 
{\color{red}{ - \! \sum_{j=1}^{n_{i}} \frac{K_{ij}}{m} \! \!
\left[ \! \! \begin{array}{ccc} \Delta_{ij}^{x} \Delta_{ij}^{x} & \Delta_{ij}^{x} \Delta_{ij}^{y} & 
\Delta_{ij}^{x} \Delta_{ij}^{z} \\ \Delta_{ij}^{y} \Delta_{ij}^{x} & \Delta_{ij}^{y} \Delta_{ij}^{y} & 
\Delta_{ij}^{y} \Delta_{ij}^{z} \\ \Delta_{ij}^{z} \Delta_{ij}^{x} & \Delta_{ij}^{z} \Delta_{ij}^{y} & 
\Delta_{ij}^{z} \Delta_{ij}^{z}  \end{array} \! \! \! \right] \! \! \! \! \left[ \! \! \begin{array}{l} \delta_{i}^{x} \\ 
\delta_{i}^{y} \\ \delta_{i}^{z} \end{array} \! \! \! \right ]}} \\ \nonumber + \! \sum_{j=1}^{n_{i}} \frac{K_{ij}}{m}  \! \!
\left[ \! \! \begin{array}{ccc} \Delta_{ij}^{x} \Delta_{ij}^{x} & \Delta_{ij}^{x} \Delta_{ij}^{y} &
\Delta_{ij}^{x} \Delta_{ij}^{z} \\ \Delta_{ij}^{y} \Delta_{ij}^{x} & \Delta_{ij}^{y} \Delta_{ij}^{y} &
\Delta_{ij}^{y} \Delta_{ij}^{z} \\ \Delta_{ij}^{z} \Delta_{ij}^{x} & \Delta_{ij}^{z} \Delta_{ij}^{y} & 
\Delta_{ij}^{z} \Delta_{ij}^{z} \end{array} \! \! \! \right ] \! \! \! \! \left[ \! \! \begin{array}{l} \delta_{j}^{x} \\ 
\delta_{j}^{y} \\ \delta_{j}^{z} \end{array} \! \! \! \right] 
\end{align}

\begin{figure}
\includegraphics[width=.49\textwidth]{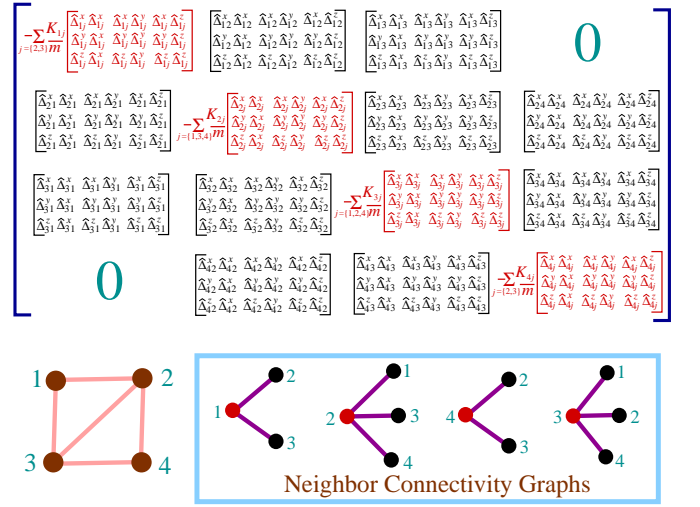}
\caption{\label{fig:Fig1} (Color online) A sample matrix is shown for a four member system depicted 
schematically below the matrix.  The connectivity graphs illustrate which neighbors correspond to 
each of the numbered particles, and the connectivity pattern dictates the form of the matrix
to be diagonalized to obtain the phonon modes.} 
\end{figure}

The phonon modes are obtained by diagonalizing the symmetric matrix, and in order to be definite, the eigenstates are taken to 
be normalized to unity.  The form of the matrix is illustrated in Fig.~\ref{fig:Fig1} for a system 
containing four particles and rendered schematically below the matrix on the far left.  Immediately to the right 
are neighbor connectivity graphs illustrating the bonding scheme which gives rise to the matrix.  The latter 
provide a way to set up the nonzero $3 \times 3$ submatrices away from the main diagonal.

In calculating thermodynamic quantities such as the mean square deviation from  
equilibrium per atom $\delta_{\mathrm{RMS}}$, one evaluates the 
partition function
\begin{align}
Z = \int (e ^{-\beta \mathcal{H}} ) d \textrm{conf}, 
\end{align} 
where $\beta = 1/k_{\mathrm{B}}$ with $k_{\mathrm{B}}$ being the Boltzmann 
constant; the integral sign and ``dconf'' refer to summing over all possible system 
configurations, where all possible atomic and velocities deviations are sampled.  
One may determine the mean square deviations from equilibrium with the aid of formalism 
developed elsewhere~\cite{tres}.  The RMS displacements from equilibrium $\delta_{\mathrm{RMS}}$, 
at the level of the harmonic approximations we have made will consist of a thermal factor 
proportional to $T^{1/2}$ and
a factor dependent on the charateristics of the lattice.  The latter be of most interest in 
this context, because it will depend on the 
rigidity associated with the specific atomic configurations and the bonding pattern 
among the atomns comprising the amorphous material.  This normalized mean square deviation 
lacks the thermal factor, and is proportional to the square root of the sum over the 
reciprocals of the squares of the phonon frequencies (with the zero frequencies excluded).
Hence, the normalized RMS displacements $\delta_{\mathrm{RMS}}^{n}$ are given by 
\begin{align}
\delta_{\mathrm{RMS}}^{n} = \left( \sum_{j=1}^{M} \omega_{j}^{-2} \right)^{1/2}
\end{align}    
where $M$ is the total number of nonzero frequencies.  
As the system becomes less rigid and zero modes proliferate, the lattice will lose local stability, 
ultimately being reduced to small isolated clusters which may move independently. 
As the integrity of the amorphous medium is lost, the notion of RMS displacements from 
positions in a single large lattice ceases to be an appropriate quantity to consider.

For a specific configuration of disorder, one 
calculates the phonon density of states by diagonalizing the appropriate matrix 
to determine the frequencies $\omega_{j}$ of the 3N phonon modes.  To build up a histogram with 
sufficient statistics, it is important to sample multiple realizations of disorder, and the 
total number of eigenvalues used in constructing histograms will be $3 \langle N \rangle n_{\mathrm{conf}}$ where 
$\langle N \rangle$ is the mean number of atoms in the unit cell, and $n_{\mathrm{conf}}$ is the total number 
of configurations sampled.  To ensure that at least $n_{\omega}$ phonon frequencies are sampled, we set 
$n_{\mathrm{conf}} = \tfrac{1}{3}(n_{\omega}/ \langle N \rangle)$.  In calculating the DOS histograms we have $n_{\omega} = 10^{6}$, 
whereas for the IPR calculations where both the 
phonon eigen-frequencies and states must be obtained, we sample 
$n_{\omega} = 10^{5}$ eigenstates.

In the highly amorphous arrangements of atoms we consider, atomic positions are uncorrelated, and hence there 
is no local order.   The probability of having $n$ atomic members in a volume $v = L^{3}$ is 
$p(n) =  \tfrac{(\rho v)^{n}}{n!}  e^{-\rho v} $ with $\rho$ being the number density.
We sample realizations of disorder in an unbiased way with a stochastic technique based on the 
Metropolis Criterion.  Beginning with the integer closest to the mean occupancy $\langle n \rangle = 
\rho v$, we make a series of attempts to change $n$ (the number of attempts is made equal to $n$ 
to ensure proper randominization) where half of the attempts would reduce $n$ by one, and 
half try to increment $n$ by one unit.  For attempts to increase $n$, the relevant parameter is 
the probability ratio $r_{+} \equiv p(n+1)/p(n) = \rho v/(n+1)$, and the increase is accepted if 
$X_{\mathrm{r}} < r_{+}$ where $X_{\mathrm{r}}$ is a random number confined to the 
interval $[0,1]$ and generated with uniform probability.  On the other hand, the appropriate probability ratio
for attempts to decrease $n$ is $r_{-} = p(n-1)/p(n) = n/\rho v$, and the occupancy is reduced by one if 
$X_{\mathrm{r}} < r{-}$ with $X_{\mathrm{r}}$ a random variable again sampled uniformly from the
interval $[0,1]$.  In generating the random numbers, a Mersenne twister algorithm is used to 
minimize correlations effects between successively generate numbers and to ensure a long period in 
the random number sequence.

With the occupancy $n$ determined, equilibrium coordinates $x_{i}^{0}$, $y_{i}^{0}$, and 
$z_{i}^{0}$ for each atom are chosen with uniform probability in the interval $[0,L]$, and the 
appropriate Hamiltonian matrix is constructed and diagonalized to calculate the photon frequencies 
to construct the vibrational DOS; the eigenstates themselves are retained to calculate the Inverse 
Participation Ratio (IPR). To reduce the impact of finite size effects, we assume periodic 
boundary conditions. 

To obtain good statistics and reach the bulk limit, we calculate the phonon frequency spectra 
for multiple disorder realizations for several different system sizes.  Averaging over disorder provides a smoother density of states curve in a 
manner which approximately mimics the DOS profile in the thermodynamic limit where some self-averaging would be 
expected to occur.  In the sequence of successively larger systems examined,
the merging of the curves indicate the attainment of the bulk limit.  

\begin{figure}
\includegraphics[width=.49\textwidth]{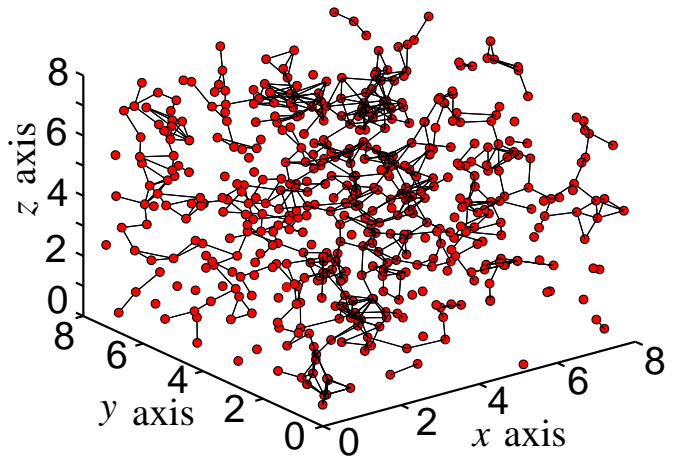}
\caption{\label{fig:Fig2} (Color online) Plot of atomic locations and bonding scheme for
$n_{\mathrm{neigh}} = 3.0$.  Red symbols represent atoms, and the dark lines connecting the spheres are bonds 
between neighbors.}
\end{figure}

\begin{figure}
\includegraphics[width=.49\textwidth]{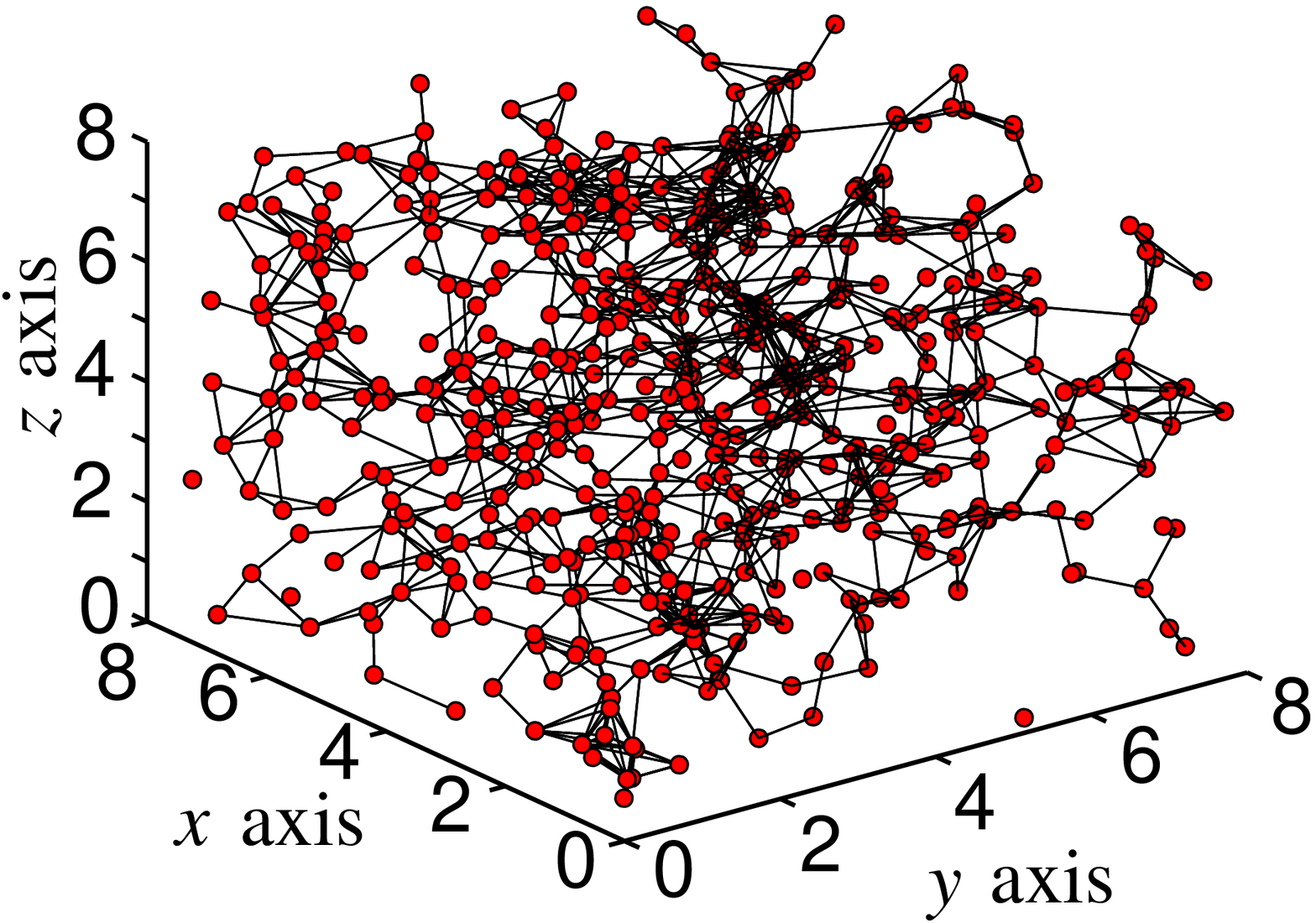}
\caption{\label{fig:Fig3} (Color online) Plot of atomic locations and bonding scheme for 
$n_{\mathrm{neigh}} = 6.0$.  Red symbols represent atoms, and the dark lines connecting the spheres are bonds 
between neighbors.}  
\end{figure}

\begin{figure}
\includegraphics[width=.49\textwidth]{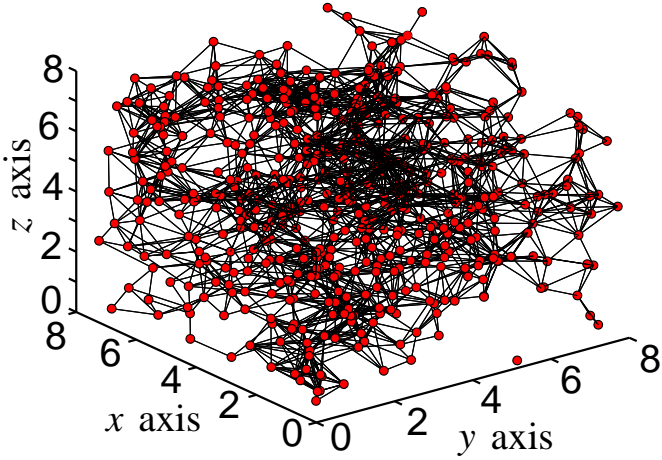}
\caption{\label{fig:Fig4} (Color online) Plot of atomic locations and bonding scheme for 
$n_{\mathrm{neigh}} = 12.0$.  Red symbols represent atoms, and the dark lines connecting the spheres are bonds
between neighbors.} 
\end{figure}

To illustrate the atomic configurations and lattice connectivity for the highly amorphous material with a 
truncated coupling between atoms, snapshots of a portion of the strongly 
disordered medium are shown in Fig.~\ref{fig:Fig2}, Fig.~\ref{fig:Fig3}, and Fig.~\ref{fig:Fig4}.  
Although the atomic configurations are the same in each case, the different values of the 
truncation radius $r_{c}$ yield very different bonding configurations.
The first image in Fig.~\ref{fig:Fig2} shows the medium with 
$n_{\mathrm{neigh}} = 3.0$ and hence a very low connectivity among the atoms.  
Fig.~\ref{fig:Fig3} 
is an intermediate case where $n_{\mathrm{neigh}} = 6.0$.  The last of the depictions in Fig.~\ref{fig:Fig4} is 
a configuration where the average number of neighbors is relatively high, and $n_{\mathrm{neigh}} = 12.0$.
The bonding scheme and its effectiveness in setting up a rigid lattice varies significantly with 
increasing $n_{\mathrm{neigh}}$.

\section{Truncated Interaction with a cutoff}

\begin{figure}
\includegraphics[width=.49\textwidth]{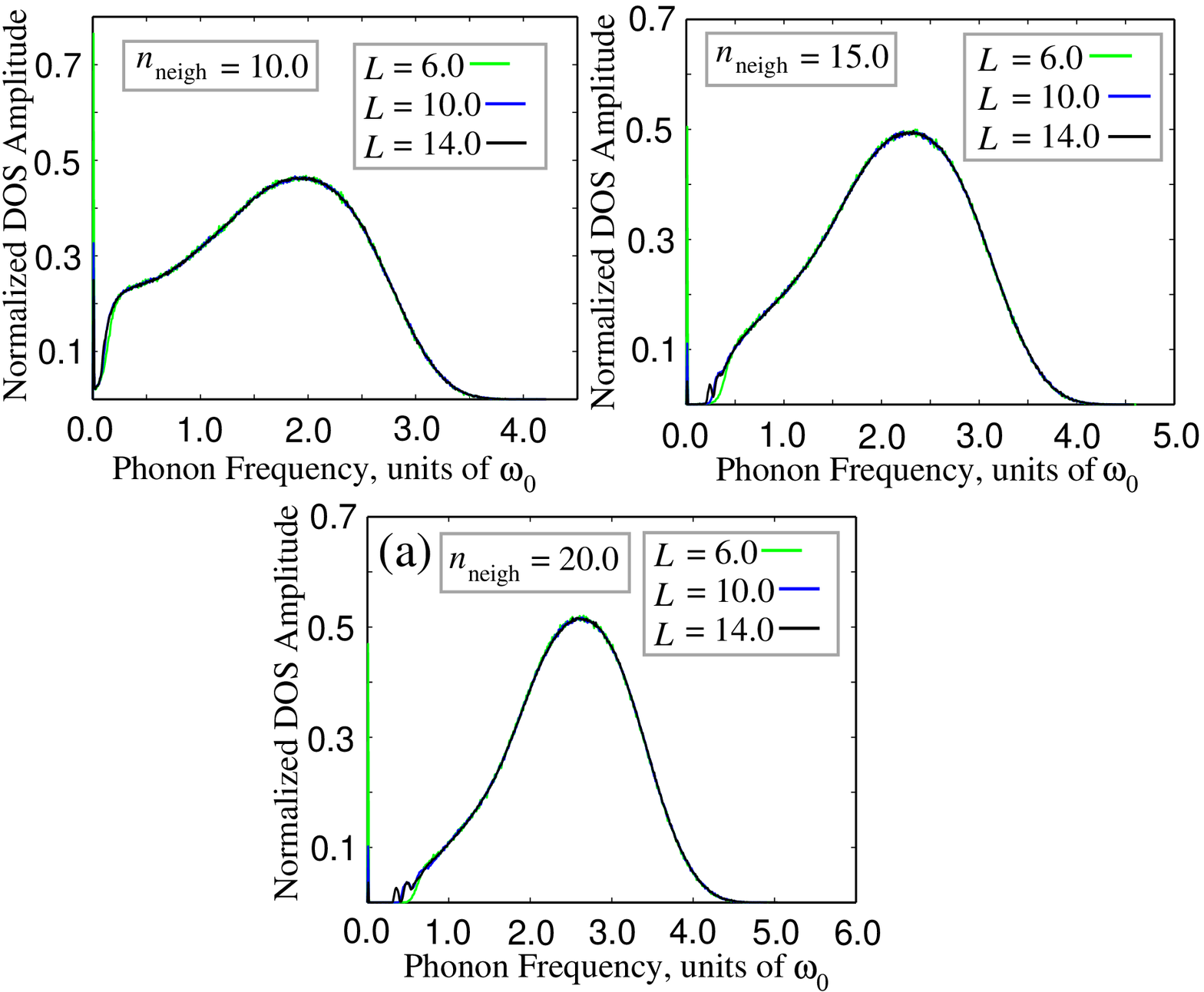}
\caption{\label{fig:Fig5} (Color online) Normalized phonon Density of States.  Panel (a) corresponds to $n_{\mathrm{neigh}} = 10.0$,
panel (b) displays DOS curves for $n_{\mathrm{neigh}} = 15.0$, and panel (c) is plotted for $n_{\mathrm{neigh}} = 20.0$.
Results for various supercell size $L$ are shown.}
\end{figure}

\begin{figure}
\includegraphics[width=.49\textwidth]{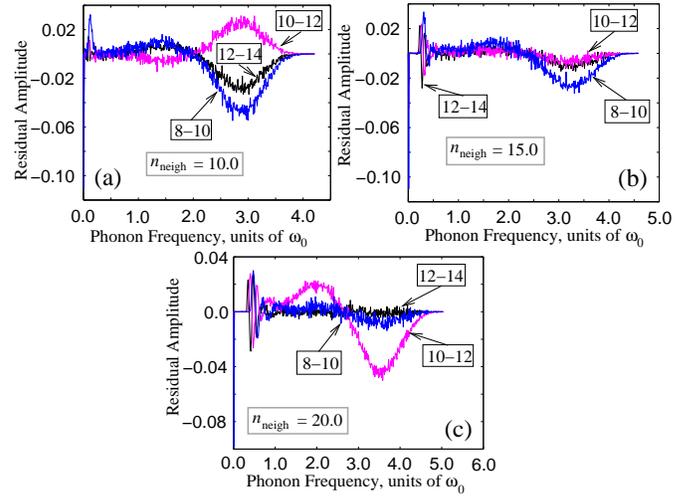}
\caption{\label{fig:Fig6} (Color online) Residual histogram amplitudes for $n_{\mathrm{neigh}} = 10.0$, $n_{\mathrm{neigh}} = 15.0$,
$n_{\mathrm{neigh}} = 20.0$ where numerals enclosed in boxes indicate the systems compared; e.g. ``12-14'' refers 
to the residual amplitude between systems with $L = 12$ and $L = 14$.}
\end{figure}

\begin{figure}
\includegraphics[width=.49\textwidth]{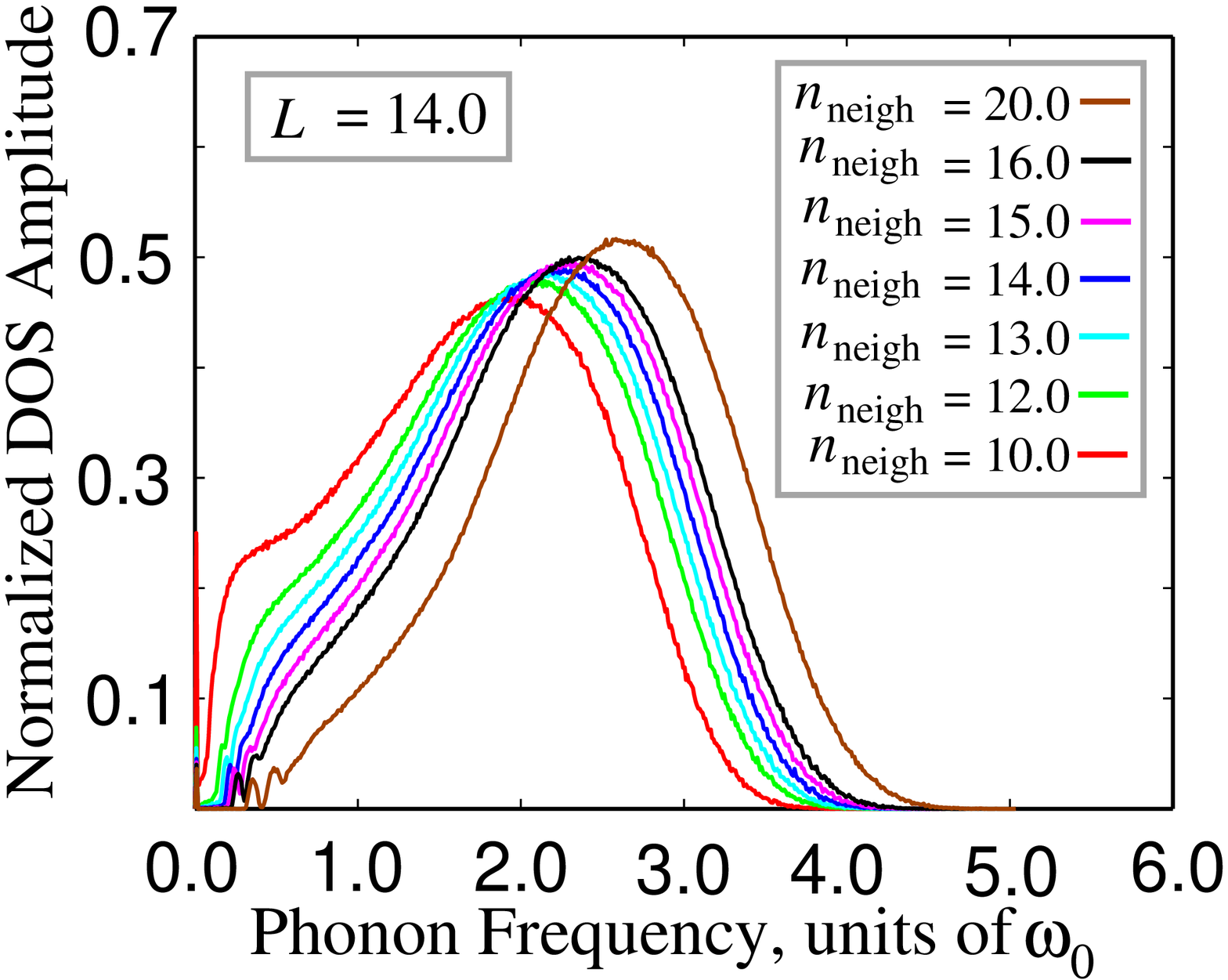}
\caption{\label{fig:Fig7} (Color online) Normalized phonon Density of States.  Panel (a) corresponds to $n_{\mathrm{neigh}} = 10.0$,
panel (b) displays DOS curves for $n_{\mathrm{neigh}} = 15.0$, and panel (c) is plotted for $n_{\mathrm{neigh}} = 20.0$.
Results for various supercell size $L$ are shown.}
\end{figure}

\begin{figure}
\includegraphics[width=.45\textwidth]{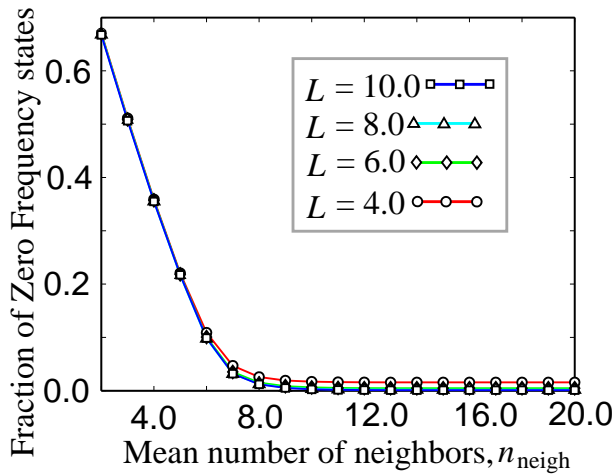}
\caption{\label{fig:Fig8} (Color online) Fraction of zero frequency phonon states for system sizes ranging from $L = 4.0$ to 
$L = 10.0$ with the mean number of neighbors, $n_{\mathrm{neigh}}$ on the horizontal axis.} 
\end{figure}

\begin{figure}
\includegraphics[width=.49\textwidth]{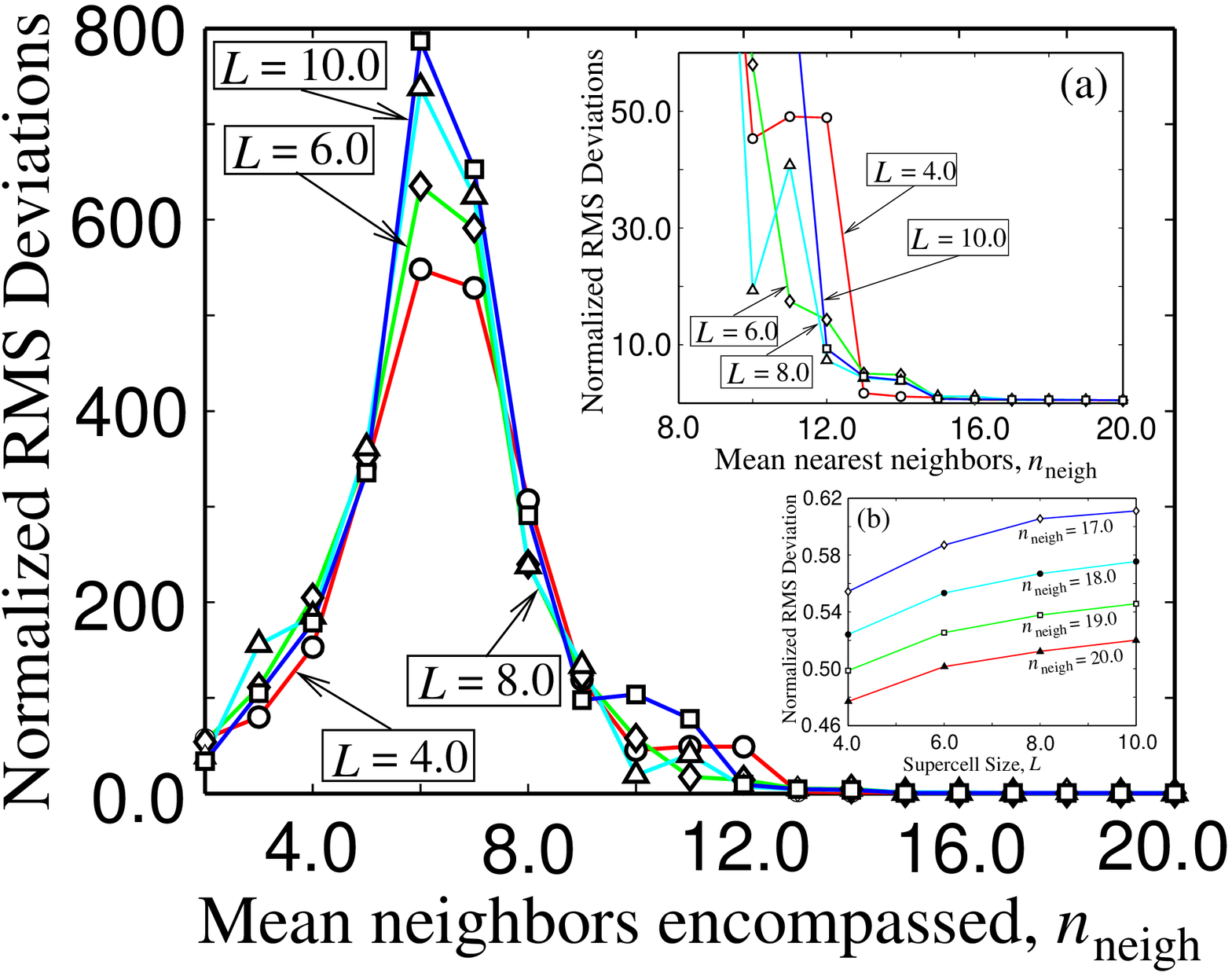}
\caption{\label{fig:Fig9} (Color online) Normalized mean square deviations plotted versus $n_{\mathrm{neigh}}$ for 
various supercell sizes $L$.  Panel (a) is a closer view of the rapid jump in the RMS fluctuations near $n_{\mathrm{neigh}} = 10.0$, and 
panel (c) show the mean square fluctuations with respect to $L$ for several relatively large values of $n_{\mathrm{neigh}}$.}
\end{figure}

We calculate the phonon density of states curves for a range of values of $r_{c}$.  As the threshold radius is 
decreased, the average number of neighbors $n_{\mathrm{neigh}}$ for each atom also 
is decreased, and it is less likely the amorphous configurations will be rigid;  
the migration of the main peak of the roughly uni-modal phonon density of states toward 
lower frequencies is compatible with this expectation.  However, one must be sure that 
$L$ is large enough that finite size effects have minimal impact on the 
DOS curves.  Figure~\ref{fig:Fig5} shows the density of states curves
for a range of system sizes, and the near coincidence of the DOS profiles for the larger systems is 
evidence of convergence with respect to the size $L$ of the supercell.  Figure~\ref{fig:Fig6} highlights the difference 
among the DOS curves and depicts the residual differences between the disorder-averaged curves
corresponding to different values of $L$.  The residual values, differences between DOS curves for    
successive supercell size $L$ shown in panels (a), (b), and (c) of Fig.~\ref{fig:Fig6} are two orders of magnitude below the principle phonon 
density of states amplitude, providing additional evidence the DOS curves are converged with respect to the 
supercell size $L$.   

The principle results are summarized in Fig.~ref{fig:Fig7}, where the DOS curves for a range of average neighbors (i.e. 
from $n_{\mathrm{neigh}} = 10.0$ to $n_{\mathrm{neigh}} = 20.0$) are graphed for the largest system size ($L = 14.0$) in 
the study.  As the average number of neighbors is decreased from $n_{\mathrm{neigh}} = 20.0$, the shape of the curve changes and the 
weight of the histogram shifts leftward, toward the lower frequency regime.  Ultimately, in the vicinity of $n_{neigh} = 10.0$, a sharp
zero frequency peak forms on the left edge of the graph, and increases in height with decreasing $r_{c}$.   

Fig.~\ref{fig:Fig8} shows the relative fraction $f$ of zero modes with respect to the mean number of nearest neighbors $n_{\mathrm{neigh}}$.
There is an abrupt increase in $f$ beginning in the vicinity of $n_{\mathrm{neigh}} = 8.0$ with an approximately linear 
increase in the zero mode fraction with decreasing $n_{\mathrm{neigh}}$ beginning for $n_{\mathrm{neigh}} < 8.0$.  The rapid proliferation of 
zero modes is a phenomenon which one may attribute to the degradation of rigidity with decreasing average number 
of nearest neighbors $n_{\mathrm{neigh}}$.  Fig.~\ref{fig:Fig9} shows the normalized (the $t^{1/2}$ factor is 
not included) root mean square (RMS) deviation which would be 
calculated thermodynamically as discussed in Section I.  Again, an abrupt (albeit noisy) increase    
in $\delta_{\mathrm{RMS}}^{n}$ begins in the vicinity of $n_{\mathrm{neigh}} = 8.0$.  Inset (a) of Fig.~\ref{fig:Fig9} 
is a closer view of the $\delta_{\mathrm{RMS}}^{n}$ 
curves near the rapid increase.  As the average number of neighbors $n_{\mathrm{neigh}}$ is decrease, the 
mean square deviation continues to increase, ultimately reaching a maximum in the vicinity of $n_{\mathrm{neigh}} = 5.0$.
The rapid decline which begins in $\delta_{\mathrm{RMS}}^{n}$ for $n_{\mathrm{neigh}} > 5.0$ is not an indication 
that the amorphous lattice has become more rigid, but is instead a result of the rapid proliferation of zero modes which 
indicate the rigidity of the medium has been compromised and the RMS deviation from equilibrium is no longer 
a reliable index of the integrity of the lattice.  

Inset (b) of Fig.~\ref{fig:Fig9} show the smoothly convergent behavior of the thermally 
excited RMS deviations for relatively large values of the average number of atomic neighbors where 
there are very few zero frequency phonon modes.
In this regime of relatively high $n_{\mathrm{neigh}}$ values, fluctuations about equilibrium 
positions in the lattice are ultimately stable with increasing $L$ and the specific configuration of atomic 
positions (albeit strongly disordered) is stable at finite temperatures in the bulk limit.

\begin{figure}
\includegraphics[width=.49\textwidth]{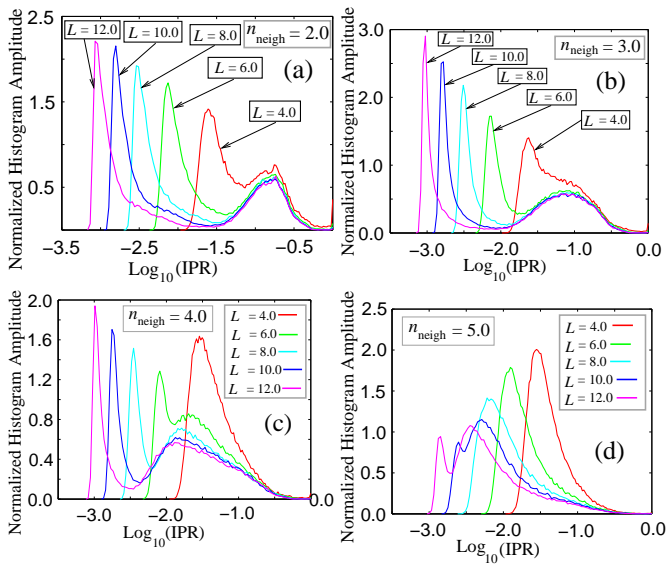}
\caption{\label{fig:Fig10} (Color online) IPR histograms for various system sizes with $n_{\mathrm{neigh}} = 2.0$ in panel (a), 
$n_{\mathrm{neigh}} = 3.0$ in panel (b), $n_{\mathrm{neigh}} = 4.0$ in panel (c), and $n_{\mathrm{neigh}} = 5.0$ in 
panel (d).}
\end{figure}

\begin{figure}
\includegraphics[width=.49\textwidth]{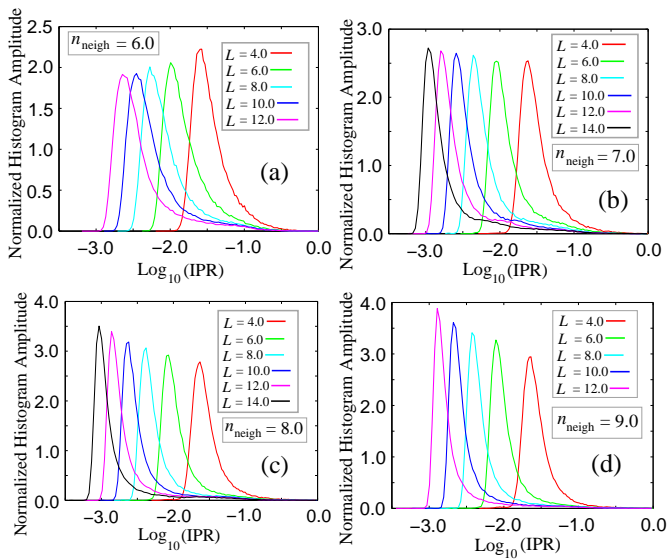}
\caption{\label{fig:Fig11} (Color online) IPR histograms for a range system sizes with $n_{\mathrm{neigh}} = 6.0$ in panel (a),
$n_{\mathrm{neigh}} = 7.0$ in panel (b), $n_{\mathrm{neigh}} = 8.0$ in panel (c), and $n_{\mathrm{neigh}} = 8.0$ in panel (d).}
\end{figure}

\begin{figure}
\includegraphics[width=.49\textwidth]{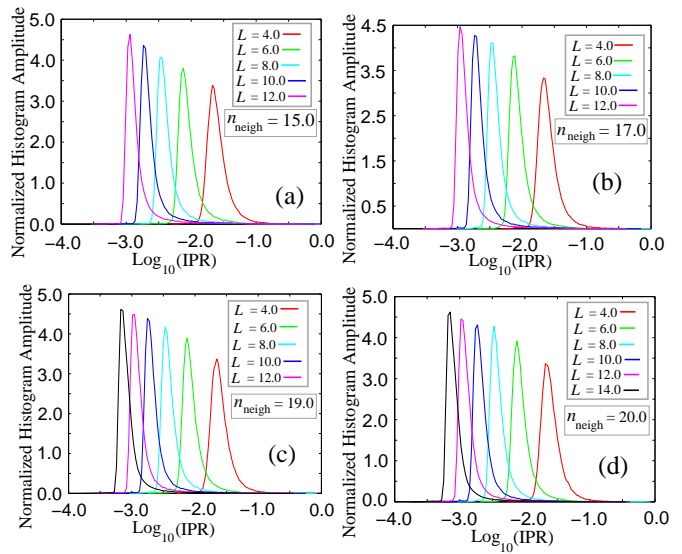}
\caption{\label{fig:Fig12} (Color online) IPR histograms for various system sizes with $n_{\mathrm{neigh}} = 15.0$ in panel (a),
$n_{\mathrm{neigh}} = 17.0$ in panel (b), $n_{\mathrm{neigh}} = 19.0$ in panel (c), and $n_{\mathrm{neigh}} = 20.0$ in panel (d).}
\end{figure}

The inverse participation ratio (IPR) is a useful index of the characteristics of phonons with respect 
to localization.  The diagnostic merit of $\kappa_{\mathrm{IPR}}$ is realized in the bulk limit where the IPR     
either converges to a finite value (for localized states) or tends to zero (for extended states).  We 
investigate the extent to which phonons are localized by calculating the probability density 
values and averaging over a sufficient number of disorder configurations (i.e. we consider at least $10^{5}$ phonon 
states).  Results are shown in Fig.~\ref{fig:Fig10},~\ref{fig:Fig11}, and ~\ref{fig:Fig12} for supercell sizes $L$ ranging from $L = 4.0$ to $L = 10.0$.  

The three sets of figures each contain four IPR histogram graphs, where $n_{\mathrm{neigh}}$ is increased 
from one panel to the next, and between successive figures.  In particular, $n_{\mathrm{neigh}}$ ranges from 
$n_{\mathrm{neigh}} = 2.0$ to $n_{\mathrm{neigh}} = 5.0$ in Fig.~\ref{fig:Fig10}, from $n_{\mathrm{neigh}} = 6.0$ 
to $n_{\mathrm{neigh}} = 9.0$ in Fig.~\ref{fig:Fig11}, and from $n_{\mathrm{neigh}} = 15.0$ to $n_{\mathrm{neigh}} = 20.0$
in Fig.~\ref{fig:Fig12} where a relatively large average number of neighbors are involved. 

The graphs in Fig.~\ref{fig:Fig10}, Fig.~\ref{fig:Fig11}, and Fig.~\ref{fig:Fig12} illustrate salient trends in the characteristics of 
the phonon states with respect to localization.  IPR curves in panels (a), (b), (c), and (d) 
of Figure~\ref{fig:Fig10} have a bimodal structure where there is a peak for relatively high IPR values as well as 
a sharp maximum at lower values of the Inverse Participation Ratio.  The two peaks 
behave very differently as the system size $L$ is increased; the portion of the histogram near the 
rightmost peak values tend to converge, ultimately shifting neither in position, weight, or 
overall form with increasing $L$.  On the other hand, the peak appearing at lower 
IPR values continues to sharpen and migrate leftward, toward even lower IPR values as $L$ 
is made larger.  
The steady shift toward lower Inverse Participation Ratios is particularly significant 
given the fact that the horizontal axis in the graph is a base ten logarithm of the IPR;
the substantial shift of density to lower IPR values with the IPR quickly shrinking  
signals the presence of extended states where  
finite size effects interfere less with the extended character of the 
corresponding phonon modes.

For the right-most peak appearing in the higher IPR regime behaves in a qualitatively 
different way with increasing $L$, eventually ceasing to move or change in shape with 
further enlargement of the supercell.  The latter indicates the presence of localized 
states where the phonon modes are characterized by a finite IPR value.  The convergence 
of the histogram curves with respect to system size may be attributed to 
the diminution of the severity of finite size effects as the regime where $L \gg \xi$ is 
attained, with $\xi$ being the localization scale of the phonon modes.  
As $n_{\mathrm{neigh}}$ is increased within the figure, the bimodal structure remains, but the 
convergence of the peak in the large IPR regime is less rapid and most of the 
statistical weight eventually shifts toward smaller IPR values, or in the direction of 
extended states. 

   In Fig.~\ref{fig:Fig11} there is a sudden change in the characteristics of the Inverse Participation 
Ratio curves where the bimodal profile gives way to a single peak structure.  The transition occurs for 
approximately the same value of $n_{\mathrm{neigh}}$ where the RMS deviation $\delta_{\mathrm{RMS}}^{n}$ 
peaks in Fig.~\ref{fig:Fig9}.  In all cases 
shown, with $n_{\mathrm{neigh}}$ ranging from $n_{\mathrm{neigh}} = 6.0$ to $n_{\mathrm{neigh}} = 9.0$, the 
peak becomes sharper and migrates toward lower values of the Inverse Participation Ratio.

Finally, results for the regime of large $n_{\mathrm{neigh}}$ are shown in Fig.~\ref{fig:Fig12}. 
In particular, panels (a), (b), (c), and (d) of Fig.~\ref{fig:Fig12} correspond to average neighbor 
values ranging from $n_{\mathrm{neigh}} = 15.0$ to $n_{\mathrm{neigh}} = 20.0$.  A salient characteristic of the 
IPR histograms is the lack of significant variation and location of the peaks for a fixed supercell size $L$.

The apparent approach of the Inverse Participation Ratio histogram curves to limiting profiles suggests that a strongly 
rigid limit has been achieved for $n_{\mathrm{neigh}} \geq 15.0$.  Statistically speaking, in this $r_{c}$ limit,
lattice vibration modes have identical characteristics with respect to localization.

\section{Exponentially Decaying Interaction}

\begin{figure}
\includegraphics[width=.49\textwidth]{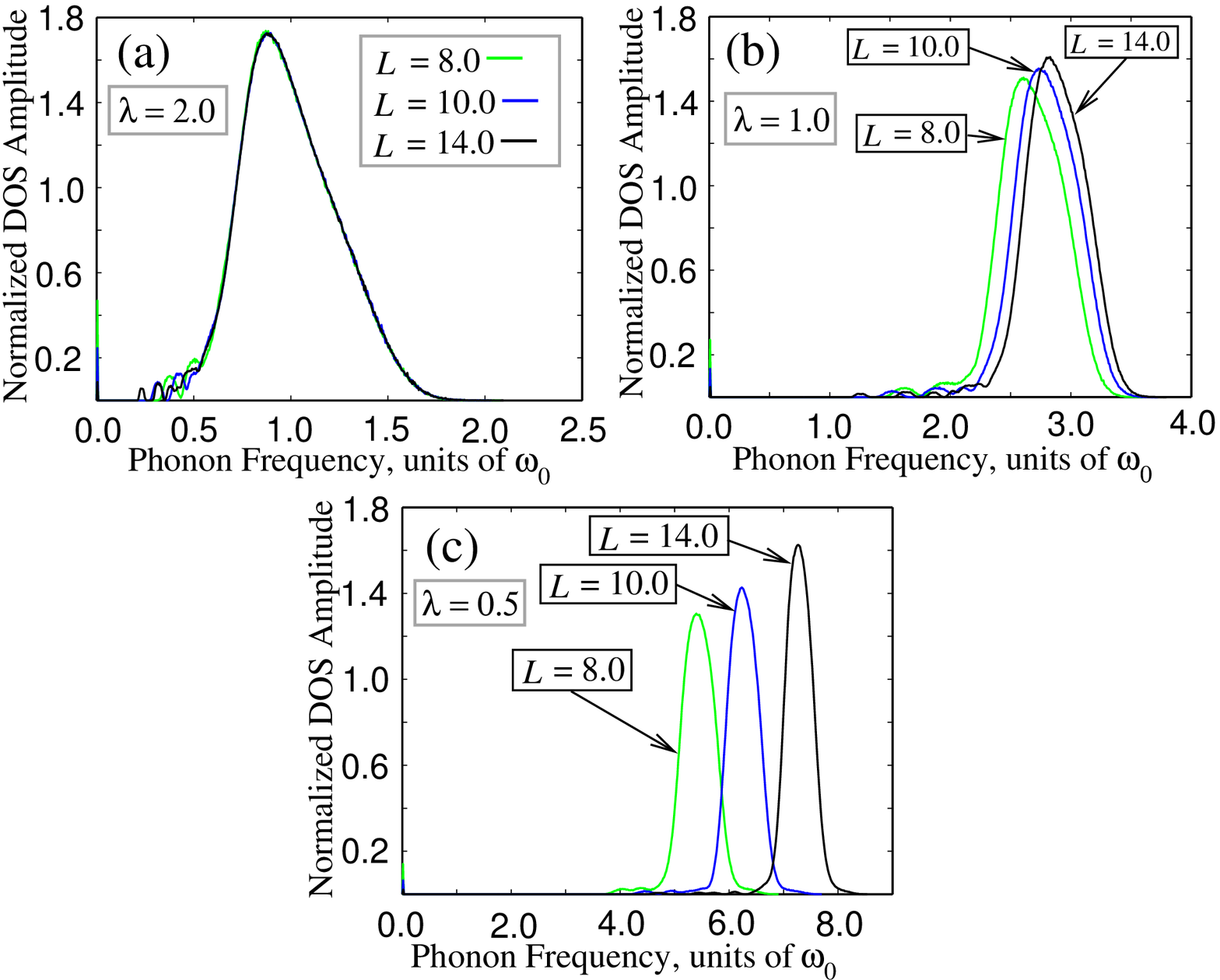}
\caption{\label{fig:Fig13} (Color Online) Normalized phonon Density of States.  
Panel (a) corresponds to $\lambda = 2.0$,
panel (b) displays DOS curves for $\lambda = 1.0$, and panel (c) is plotted for $\lambda = 0.5$.
Results for various supercell size $L$ are shown.}
\end{figure}

An exponentially decaying interaction scheme provides a short-ranged interaction without the 
need for an abrupt cutoff $r_{c}$ between atoms is taken to be proportional to $e^{-\lambda r}$ where 
$1/\lambda$ provides a length scale.
The DOS curves shown in Fig.~\ref{fig:Fig13} are obtained for a range of decay constants $\lambda$.  Panels (a), (b), and (c)
of Fig.~\ref{fig:Fig13} correspond to decay constants $\lambda = 2.0$, $\lambda = 1.0$, and $\lambda = 0.50$.  In 
seeking the bulk limit, two relevant length scales are the typical separation $l = \rho^{-1/3}$ 
(unity in the present case where $\rho = 1$) between atoms and $\lambda^{-1}$ for the 
interatomic potential.  Hence, one must have $L \gg \max \left \{ \rho^{-1/3}, \lambda^{-1} \right \}$ 
for convergence to the bulk limit.  

The relatively rapid convergence of the phonon DOS curves for $\lambda = 2.0$ is compatible 
with the short range of the interaction between atoms.  For the relatively long-range cases $\lambda = 1.0$ and  
$\lambda = 0.50$, convergence is less rapid with the slowest approach to the bulk limit occurring in panel (c) of 
Fig.~\ref{fig:Fig13}, and a slightly more rapid attainment of the thermodynamic limit in the intermediate case displayed in 
panel (b).

\begin{figure}
\includegraphics[width=.49\textwidth]{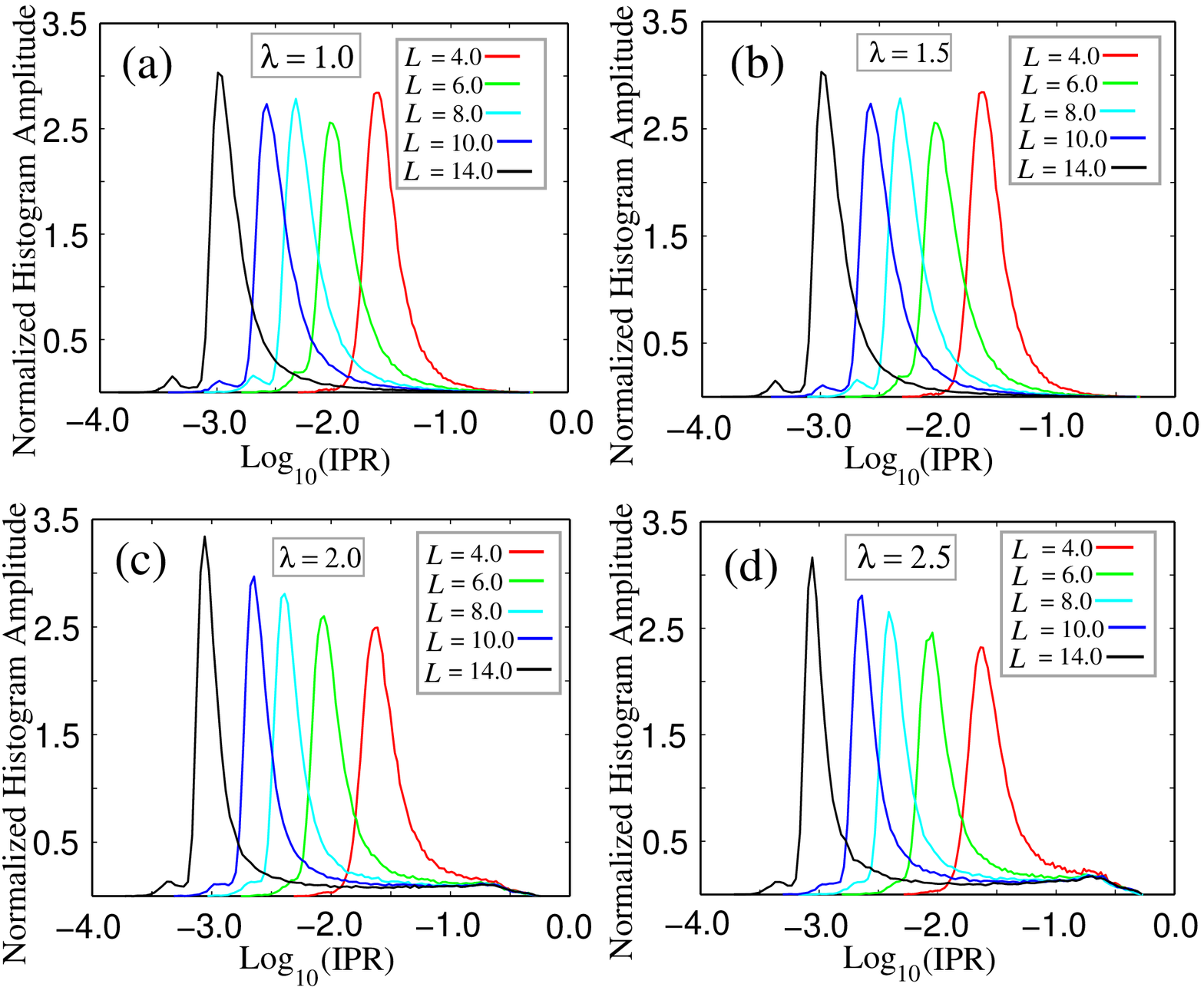}
\caption{\label{fig:Fig14} (Color Online) Normalized phonon Density of States.
Panel (a) corresponds to $\lambda = 2.0$,
panel (b) displays DOS curves for $lambda = 1.0$, and panel (c) is plotted for $\lambda = 0.5$.
Results for various supercell size $L$ are shown.}
\end{figure}

As in the case of the interaction with a cutoff scale $r_{c}$, we calculate IPR histograms for the exponentially
decaying inter-atomic coupling, with results appearing in Fig.~\ref{fig:Fig14}.  
The changes in the characteristics of the states with respect to localization are less dramatic 
with respect to localization with changes in $\lambda^{-1}$ than with different $r_{c}$ values 
for the case of the truncated potential.  In panels (a), (b), (c), and (d) of Fig.~\ref{fig:Fig14}, the 
IPR histogram is shown for coupling decay constants $\lambda = 1.0$, $\lambda = 1.5$, $\lambda = 2.0$, and $\lambda = 2.5$. 
If the interaction is relatively long-ranged (i.e. 
comparable to or greater than the interatomic separation $l = \rho^{-1/3}$), very little change occurs in 
the shapes of the histogram curves for a particular system size $L$.  The IPR histogram profiles are 
uni-modal, and the tendency of the peaks to become narrower and shift in the direction of lower IPR values 
is similar to what is seen for the IPR histograms corresponding to the truncated potential in the limit that 
the average number of neighbors is large.  With shorter ranged and more rapidly decaying 
couplings between atoms, changes in the IPR histograms are still relatively subtle, in 
contrast to the abrupt shifts occurring in the case of the sharply truncated potential as 
$r_{c}$ is reduced.  A feature which does appear in the case of the more rapidly decaying 
potentials [e.g. for $\lambda = 2.0$ shown in panel (c) of Fig.~\ref{fig:Fig14} and $\lambda = 2.5$ in 
panel (d) of Fig.~\ref{fig:Fig14}], is a small shoulder on the right side of the graph for relatively 
large values of the Inverse Participation Ratio.  Although the structure is relatively weak, it persists at approximately the 
same amplitude with increasing $L$, suggesting some of the phonon states are at least quasi-localized 
if the length scale $\lambda^{-1}$ of the exponential coupling is sufficiently small.
This behavior is qualitatively similar, though markedly subtler, than the convergence of the
right-most peak corresponding to localized states for the case of the truncated potential
where $n_{\mathrm{neigh}} < 5.0$.  

\section{Conclusions}
We have examined characteristics of phonon states in strongly disordered media where the coupling between 
atoms is finite in range, with either a flat profile terminating at a radius $r_{c}$, or a more gradually  
decaying exponential coupling scheme with its own length scale $\lambda^{-1}$.  
In the case of the abruptly truncated interaction, an interatomic potential profile that may be more appropriate for relatively loosely packed
particles with a non-covalent more rapidly decaying Lennard-Jones type of coupling, there are 
sudden transitions in shapes of the phonon DOS curves and the root mean square deviations $\delta_{\mathrm{RMS}}$ 
from equilibrium with decreasing $r_{c}$.  In addition, the Inverse Participation Ratios change qualitatively between 
$n_{\mathrm{neigh}} = 5.0$ and $n_{\mathrm{neigh}} = 6.0$, converting from a bimodal profile 
with localization signatures to a single peak which migrates to lower IPR values with increasing 
supercell size as expected for extended states.

For the exponential coupling profile, which decreases gradually with the 
inter-atomic distance, the variation of the IPR histogram with decreasing 
range $\lambda^{-1}$ is not punctuated by sharp changes. Moreover, significant fraction of the phonon states have 
extended character even for the relatively short-ranged case $\lambda = 2.5$.

\begin{acknowledgments}
\end{acknowledgments}


\end{document}